\documentclass{amsart}
\usepackage{amsfonts}
\usepackage{amsmath}
\usepackage{amssymb}

\theoremstyle{plain}
\newtheorem{theorem}{Theorem}[section]
\newtheorem{lemma}[theorem]{Lemma}

\theoremstyle{definition}

\theoremstyle{remark}
\newtheorem{problem}[theorem]{Problem}
\newtheorem{remark}[theorem]{Remark}
\numberwithin{equation}{section}

\input{tcilatex}

\begin{document}
\title{Differential-algebraic approach to constructing representations of
commuting differentiations in functional spaces and its application to
nonlinear integrable dynamical systems }
\author{Anatolij K. Prykarpatski}
\address{The Department of Applied Mathematics at AGH University of Science
and Technology of Krakow, Poland}
\email{pryk.anat@ua.fm, prykanat@cybergal.com}
\author{ Kamal N. Soltanov}
\email{sultan\_kamal@hotmail.com, soltanov@hacettepe.edu.tr}
\author{Emin \"{O}z\c{c}a\u{g}}
\email{ozcag1@hacettepe.edu.tr}
\address{The Department of Mathematics at the Hacettepe University of
Ankara, Turkey}
\maketitle

\begin{abstract}
There is developed a differential-algebraic approach to studying the
representations of commuting differentiations in functional differential
rings under nonlinear differential constraints. An example of the
differential ideal with the only one conserved quantity is analyzed in
detail, the corresponding Lax type representations of differentiations are
constructed for an infinite hierarchy of nonlinear dynamical systems of the
Burgers and Korteweg-de Vries type. A related infinite bi-Hamiltonian
hierarchy of Lax type dynamical systems is constructed.
\end{abstract}

\section{\protect\bigskip \protect\bigskip Introduction}

We consider the ring $\mathcal{K}:=\mathbb{R}\{\{x,t\}\},$ $(x,t)\in \mathbb{%
R}^{2},$ of convergent germs of real-valued smooth Schwartz type functions
from $S(\mathbb{R}^{2};\mathbb{R})$ and construct the associated
differential quotient ring $\mathcal{K}\{u\}:=Quot(\mathcal{K}[\Theta u])$
with respect to a functional variable $u\in \mathcal{K},$ where $\Theta $
denotes \cite{Ka,Ri,GD0,GD,GMD} the standard monoid of all commuting
differentiations $D_{x}$ and $D_{t},$ satisfying the standard Leibnitz rule,
and defined by the natural conditions 
\begin{equation}
D_{x}(x)=1=D_{t}(t),\ \ \ D_{t}(x)=0=D_{x}(t),  \label{R00}
\end{equation}%
The ideal $I\{u\}\subset \mathcal{K}\{u\}$ is called differential if the
condition $I\{u\}=\Theta I\{u\}$ holds. In the differential ring $\mathcal{K}%
\{u\}$ the differentiations 
\begin{equation}
D_{t},\ D_{x}:\ \mathcal{K}\{u\}\rightarrow \mathcal{K}\{u\},  \label{R0}
\end{equation}%
satisfy the algebraic commuting relationship%
\begin{equation}
\lbrack D_{t},\ D_{x}]=0.  \label{R1}
\end{equation}%
For an arbitrarily chosen function $u\in \mathcal{K}\ $ the only
representation of (\ref{R1}) in the $\mathcal{K}\{u\}$ is of the form 
\begin{equation}
D_{t}=\partial /\partial t,\text{ \ \ \ \ }D_{x}=\partial /\partial t,
\label{R2}
\end{equation}%
being the usual partial differentiations. Nonetheless, if the function $u\in 
\mathcal{K}\ $satisfies some additional \ nonlinear differential-algebraic
constraint $Z[u]:=Z(u,D_{t}u,D_{x}u,...)$ $=0,\ $ imposed on the ring $%
\mathcal{K}\{u\}$ for some element $Z[u]\in \mathcal{K}\{u\},$ other
nontrivial representations of the differentiations (\ref{R0}) in the
corresponding reduced and invariant differential ring $\mathcal{\overline{K}}%
\{u\}:=\mathcal{K}\{u\}|_{Z[u]=0}$ $\subset \mathcal{K}\{u\}$ can exist.

Below we will consider in detail this situation and construct the
corresponding representations of the commuting relationship (\ref{R1}),
which are polynomially dependent on $u\in \mathcal{K}\ $and its derivatives
with respect to the differentiation $D_{x}.$ The found representations of
commuting differentiations $D_{t}$ and $D_{x}\ $are interpreted as the
corresponding Lax type representations for an infinite hierarchy of
nonlinear dynamical systems of Burgers and Korteweg de Vries type.

\begin{remark}
There are interesting applications \cite{PAPP,BPS} of the problem above in
the case when the differentiations $D_{t}$ and $D_{x}:\mathcal{K}%
\{u\}\rightarrow \mathcal{K}\{u\}$ are defined by means of the Lie algebraic
conditions%
\begin{equation}
\lbrack D_{x},D_{t}]=(D_{x}u)D_{x}  \label{R1a}
\end{equation}%
and 
\begin{equation}
D_{x}(x)=1=D_{t}(t),\ \ \ D_{t}(x)=u,\text{ \ }D_{x}(t)=0  \label{R00a}
\end{equation}%
for $(x,t)\in \mathbb{R}^{2}$ and $u\in \mathcal{K}.$
\end{remark}

The corresponding representations of the Lie algebraic relationships (\ref%
{R1a}) and (\ref{R00a}) under the differential constraints $%
D_{t}^{N+1}u=0,N\in {\mathbb{Z_{+}}},$ of Riemann type imposed on the ideal\
\ $\mathcal{K}\{u\},$ \ prove to be finite dimensional, being equivalent to
their so called Lax type representations, important for the integrability
theory \cite{Bl,No,PM,Wi,BPS} of nonlinear hydrodynamical systems on
functional manifolds.

\section{Differential rings with the only conserved quantity constraint}

Let us pose the following problem:

\begin{problem}
To describe the possible representations of the differentiations $D_{t},\
D_{x},$ in the ring $\mathcal{\bar{K}\{}u\mathcal{\}=K}\{u\}|_{Z[u]=0},%
\mathcal{\ }$ defined by a constraint $Z[u]\in \mathcal{K\{}u\},$ $u\in 
\mathcal{K},$ and satisfying the conditions (\ref{R00}) and (\ref{R1}).
\end{problem}

As is easy to observe, for the case of arbitrarily chosen function $u\in 
\mathcal{K}$ \ and the representation of (\ref{R1}) in the ring $\ \ 
\mathcal{K\{}u\}$ is given by the unique expressions (\ref{R2}). Another
situation arises if there is some differential constraint $%
Z[u]:=Z(u,D_{x}u,D_{t}u,...)=0,$ $Z[u]\in \mathcal{K\{}u\},$ imposed on the
function $u\in \mathcal{K}.$ Then one can expect \ that the commutation
condition (\ref{R1}), if realized in the constrained ideal $\mathcal{%
\overline{K}\{}u\}:\mathcal{=K\{}u\mathcal{\}}|_{Z[u]\ =0},$ will be much
more specified and in some cases the corresponding representations may
appear to be even finite dimensional. The latter may by of nontrivial
interest for some applications in applied sciences, especially when these
imposed constraints possess some interesting physical interpretation. To be
further more precise, we need to involve here some additional
differential-algebraic preliminaries \cite{GD,GD1,GD2,GMD}.

Consider the ring $\mathcal{K\{}u\},$ $u\in \mathcal{K},$ and the exterior
differentiation $d:\mathcal{K\{}u\}\rightarrow \Lambda ^{1}(\mathcal{K\{}%
u\}),$..., $d:\Lambda ^{p}(\mathcal{K\{}u\})\rightarrow \Lambda ^{p+1}(%
\mathcal{K\{}u\})\ $for $p\in \mathbb{Z}_{+},$ acting in the freely
generated Grassmann algebras $\Lambda (\mathcal{K\{}u\})=\oplus _{p\in 
\mathbb{Z}_{+}}\Lambda ^{p}(\mathcal{K\{}u\}),$ where by definition,%
\begin{eqnarray}
\Lambda ^{1}(\mathcal{K\{}u\}) &:&=\ \ \mathcal{K\{}u\mathcal{\}}dx+\ 
\mathcal{K\{}u\}dt+\sum_{j,k\in \mathbb{Z}_{+}}\mathcal{K\{}u\mathcal{\}}%
du^{(j,k)},\text{ \ \ }u^{(j,k)}:=D_{t}^{j}D_{x}^{k}u,  \label{R5} \\
\Lambda ^{2}(\mathcal{K\{}u\}) &:&=\mathcal{K\{}u\mathcal{\}}d\Lambda ^{1}(%
\mathcal{K\{}u\}),...,\text{ \ \ }\Lambda ^{p+1}(\mathcal{K\{}u\}):=\mathcal{%
K\{}u\mathcal{\}}d\Lambda ^{p}(\mathcal{K\{}u\}),  \notag
\end{eqnarray}%
The triple $\mathcal{A}:\mathcal{=}(\mathcal{K\{}u\},\Lambda (\mathcal{K\{}%
u\}),d)$ is called \textit{the Grassmann differential algebra} \cite{GMD}
with generatrix $u\in \mathcal{K}.$ In the algebra $\mathcal{A}$ one
naturally defines the action of differentiations $D_{t},D_{x}$ and $\
\partial /\partial u^{(j,k)}:\mathcal{A}\rightarrow \mathcal{A},j,k\in 
\mathbb{Z}_{+},$ as follows: 
\begin{eqnarray}
D_{t}u^{(j,k)} &=&u^{(j+1,k)},D_{x}u^{(j,k)}=u^{(j,k+1)},  \label{R5a} \\
D_{t}du^{(j,k)} &=&du^{(j+1,k)},D_{x}du^{(j,k)}=du^{(j,k+1)}  \notag \\
dP[u] &=&\sum_{j,k\in \mathbb{Z}_{+}}\ (\pm )\partial P[u]/\partial
u^{(j,k)}\wedge du^{(j,k)}:=P^{\prime }[u]\wedge du,  \notag
\end{eqnarray}%
where the sign $"\wedge "$ denotes the standard \cite{Go} exterior
multiplication in the differential Grassmann algebra $\Lambda (\mathcal{K\{}%
u\}),$ and for any $P\in \Lambda (\mathcal{K\{}u\})$ the mapping 
\begin{equation}
P^{\prime }[u]\wedge :\Lambda (\mathcal{K\{}u\})\rightarrow \Lambda (%
\mathcal{K\{}u\})  \label{R5aa}
\end{equation}%
\ \ is a linear differential operator in $\Lambda (\mathcal{K\{}u\}).$ The
following commutation properties 
\begin{equation}
D_{x}d=d\text{ }D_{x},\text{ \ \ \ }D_{t}d=d\text{ }D_{t}  \label{R5aaa}
\end{equation}%
hold in the Grassmann differential algebra $\mathcal{A}.$ \ The following
remark \cite{GMD} is also important.

\begin{remark}
The Lie derivative $L_{V}:\mathcal{K\{}u\}\mathcal{\rightarrow K\{}u\}\ $%
with respect to a vector field $V:\mathcal{K\{}u\}\rightarrow T(\mathcal{K\{}%
u\}),$ satisfying the condition $\ L_{V}:\mathcal{K}\mathcal{\subset }%
\mathcal{K},$ can be uniquely extended to the differentiation $L_{V}:%
\mathcal{A}\rightarrow \mathcal{A},$ satisfying the commutation condition $%
L_{V}d=d$ $L_{V}.$
\end{remark}

The \textit{variational derivative, or the functional gradient }$\nabla
P[u]\wedge \in \Lambda (\mathcal{K\{}u\}),$ \textit{is defined }for any $%
P\in \Lambda (\mathcal{K\{}u\})$ \textit{by means of the following
expression: }%
\begin{equation}
\nabla P[u]\wedge :=\sum_{j,k\in \mathbb{Z}_{+}}(-1)^{\pm
}(-D_{t})^{j}(-D_{x})^{k}\ (\partial P[u]/\partial u^{(j,k)})\wedge ,
\label{R5b}
\end{equation}%
which can be equivalently rewritten as 
\begin{equation}
\nabla P[u]\wedge =P^{\prime }[u]^{\ast }\wedge ,  \label{R5c}
\end{equation}%
where a mapping $P^{\prime }[u]^{\ast }\wedge :\Gamma (\mathcal{K\{}%
u\})\rightarrow \Gamma (\mathcal{K\{}u\})$ is formally adjoint to that of \ (%
\ref{R5aa}) and is naturally defined on the space $\Gamma (\mathcal{K\{}u\})$
of vector fields on $\mathcal{K\{}u\}.$ The latter is strongly based on the
following important lemma, stated for a special case in \cite%
{GMD,GD0,GD,GD1,GD2,Ol}.

\begin{lemma}
\label{Lm_R3.1}Let the differentiations $D_{x}$ and $D_{t}:\Lambda (\mathcal{%
K\{}u\})$ $\rightarrow \Lambda (\mathcal{K\{}u\})$ satisfy the conditions \ (%
\ref{R5a}). Then the relation%
\begin{equation}
\begin{array}{c}
Ker\nabla /(\mathrm{Im}d\oplus \mathbb{R)\simeq }H^{1}(\Lambda (\mathcal{K\{}%
u\})):= \\ 
=Ker\{d:\Lambda ^{1}(\mathcal{K\{}u\})\rightarrow \Lambda ^{2}(\mathcal{K\{}%
u\})\}/d\Lambda ^{0}(\mathcal{K\{}u\})%
\end{array}
\label{R5d}
\end{equation}
is a canonical isomorphism, where $H^{1}(\mathcal{A})\ $ is the
corresponding cohomology classes of the Grassmann complex $\Lambda (\mathcal{%
K\{}u\}).$
\end{lemma}

It is well known that in the case of the differential ring $\mathcal{K\{}u\}$
all of the cohomology classes $H^{j}(\mathcal{A}),j\in \mathbb{Z}_{+},$ are
trivial, \ giving rise to the well known classical relationship $Ker\nabla =%
\mathrm{Im}D_{x}\oplus \mathrm{Im}D_{t}\oplus \mathbb{R}.$ In addition, the
following simple relationship holds:%
\begin{equation}
\nabla \cdot (\mathrm{Im}D_{x}\oplus \mathrm{Im}D_{t})=0.  \label{R5.e}
\end{equation}

Based on Lemma \ \ref{Lm_R3.1} one can \ define the equivalence class $%
\widetilde{\mathcal{A}}:=\mathcal{A}/\{\mathrm{Im}D_{x}\oplus \mathrm{Im}%
D_{t}\oplus \mathbb{R\}}$ $:\mathbb{=}\mathcal{D(A};dxdt),$ whose elements
will be called \textit{functionals, }that is any element $\gamma \in $ $%
\mathcal{D(A};dxdt)$ can be\textit{\ } represented as a suitably defined
integral $\gamma :=\int \int dxdt\gamma \lbrack u]\in $ $\mathcal{D(A};dxdt)$
for some $\gamma \lbrack u]\in \Lambda (\mathcal{K\{}u\})$ with respect to
the Lebesgue measure $dxdt$ on $\mathbb{R}^{2}.$

Proceed now to treating the case when the following simple enough and
uniform in the variables $(x,t)\in \mathbb{R}^{2}$ differential constraint
is imposed on a function $u\in \mathcal{K}$ generating the ideal $\mathcal{%
K\{}u\}:$%
\begin{equation}
D_{t}u+K[u]=0,  \label{R3}
\end{equation}%
where a smooth element $K[u]:=D_{x}k[u]$ for some element $\ \ \ k[u]\in 
\mathcal{\overline{K}\{}u\mathcal{\}}$ is defined by means of the following
differential-algebraic condition: \textit{except the relationship (\ref{R3}%
), that is }%
\begin{equation}
D_{t}u+D_{x}k[u]=0,  \label{R3a}
\end{equation}%
\textit{there exists no other additional constraint of the form }%
\begin{equation}
D_{t}h[u]+D_{x}H[u]=0  \label{R4}
\end{equation}%
\textit{for some elements }$h[u],H[u]\in \overline{\mathcal{A}}$ \ \textit{%
from the reduced Grassmann differential algebra }$\overline{\mathcal{A}}:=(%
\bar{K}\{u\},\Lambda (\bar{K}\{u\}),d).$

\begin{remark}
It is necessary to mention that additional conservation relationships like (%
\ref{R4}) can exist in an extended differential algebra $\widetilde{\mathcal{%
A}}:=(\tilde{K}\{u\},\Lambda (\tilde{K}\{u\}),d),$ where $\tilde{K}\{u\}:=%
\bar{K}\{u;\alpha \}$ for some additional set of elements $\alpha \subset 
\mathcal{K}.$
\end{remark}

Based on the differential-algebraic setting, described above, the latter
condition can be interpreted naturally in the following way: if to define
the \ spaces of functionals $\mathcal{D(}\overline{\mathcal{A}};dx):=$ $%
\overline{\mathcal{A}}/D_{x}\overline{\mathcal{A}}$ and $\mathcal{D(}%
\overline{\mathcal{A}};dt)=$ $\overline{\mathcal{A}}/D_{t}\overline{\mathcal{%
A}}$ on the \textit{reduced Grassmann differential algebra }$\overline{%
\mathcal{A}},$ then from the functional point of view these \ factor spaces $%
\mathcal{D(\overline{\mathcal{A}}};dx)$ and $\mathcal{D(\overline{\mathcal{A}%
}};dt)$ can be understood more classically as the corresponding spaces of
suitably defined integral expressions subject to the measures $dx$ and $dt,$
respectively. Then the relationship (\ref{R4}) means equivalently that the
functional $h:=\int dxh[u]\in \mathcal{D(}\overline{\mathcal{A}};dx)$ is a
conserved quantity for the differentiation $D_{t},$ and the functional $%
H:=\int dtH[u]\in \mathcal{D(}\overline{\mathcal{A}};dt)$ is a conserved
quantity for the differentiation $D_{x}.$

Since the differential relationship \ (\ref{R3}) naturally defines on the
reduced ring $\mathcal{\overline{K}\{}u\mathcal{\}}$ the vector field $K:%
\mathcal{\overline{K}\{}u\mathcal{\}\rightarrow }T\mathcal{(\overline{K}\{}u%
\mathcal{\})},$ one can construct the corresponding Lie derivative $L_{K}:%
\overline{\mathcal{A}}\rightarrow \overline{\mathcal{A}}$ along this vector
field and calculate the quantity $L_{K}\varphi \lbrack u]\in \Lambda (%
\mathcal{\overline{K}\{}u\mathcal{\}}),$ $\varphi \lbrack u]:=\nabla \gamma
\lbrack u]\wedge \in \Lambda (\mathcal{\overline{K}\{}u\mathcal{\}})$ for
any conserved quantity $\gamma \in \mathcal{D(}\overline{\mathcal{A}};dx)$
with respect to the differentiation $D_{t}.$ The following classical lemma 
\cite{La,Ol,BPS,PM,Ol} holds.

\begin{lemma}
\label{Lm_R3.2}\textbf{\ (\textit{E. Noether, P.Lax}) }Let a quantity $%
\varphi \lbrack u]du\in \Lambda ^{1}(\mathcal{\overline{K}\{}u\mathcal{\}})$
be such that the following conditions:%
\begin{equation}
D_{t}\varphi \lbrack u]+K^{\prime ,\ast }[u]\varphi \lbrack u]=0,\varphi
^{\prime ,\ast }[u]=\varphi ^{\prime }[u],  \label{R6}
\end{equation}%
hold for any $u\in \mathcal{K},$ satisfying the differential constraint \ (%
\ref{R3}). Then the functional 
\begin{equation}
\gamma :=\int_{0}^{1}d\lambda \int dx\varphi \lbrack \lambda u]u\in \mathcal{%
D(}\overline{\mathcal{A}};dx)  \label{R7}
\end{equation}%
is a conserved quantity with respect to the differentiation $D_{t}.$
\end{lemma}

As a simple corollary from the Lemma \ \ref{Lm_R3.2}, subject to the problem
posed above, we need to show that the only solution to the equation (\ref{R6}%
) is the trivial element $\varphi \lbrack u]\ =1\in \Lambda ^{0}(\mathcal{%
\overline{K}\{}u\mathcal{\}}).$ To solve this problem effectively enough
(but not in a most general form) we will take into account \cite{CL,Sh} that
the linear equation (\ref{R6}) possesses a partial asymptotical solution
with respect to the parameter $\lambda \in \mathbb{R}$ as $|\lambda
|\rightarrow \infty ,$ such that the element 
\begin{equation}
\varphi \lbrack u]\ \sim \psi \lbrack u;\lambda ]\ \exp \{-\lambda
^{n}t-\lambda x+\partial ^{-1}u\}  \label{R8}
\end{equation}%
for any $n\in \mathbb{Z}_{+}\backslash \{0,1\}$ belongs to the finally
extended differential algebra $\Lambda ^{0}(\mathcal{\tilde{K}\{}u\mathcal{\}%
})\simeq $ $\Lambda ^{0}(\mathcal{\overline{K}\{}u;\partial ^{-1}u\mathcal{\}%
})\ $by means of the element $\partial ^{-1}u\in \mathcal{K}.$ Here, by
definition, $D_{x}\cdot \partial ^{-1}=\mathbf{1}:$ $\Lambda (\mathcal{%
\overline{K}\{}u\mathcal{\}})\rightarrow \Lambda (\mathcal{\overline{K}\{}u%
\mathcal{\}})$ is the identity mapping and, by construction, \ the element $%
\psi \lbrack u;\lambda ]\ \in \Lambda ^{0}(\mathcal{\overline{K}\{}u\mathcal{%
\}})$ and satisfies the limiting condition $\lim_{|\lambda |\rightarrow
\infty }\psi \lbrack u;\lambda ]:=1.$ It is easy to observe that the
expression (\ref{R8}) guarantees that the only conservative quantity with
respect to the differentiation $D_{t}$ is the functional $\gamma _{0}=\int
dxu\in \mathcal{D(}\overline{\mathcal{A}};dx).$ Having taken into account
the condition that the expression $\varphi \lbrack u;\lambda ]\ =1$ is the
only solution to the equation (\ref{R6}) for all $u\in \mathcal{K},$
satisfying the constraint (\ref{R3}), one easily finds from (\ref{R8}) that 
\begin{equation}
u=\lambda -D_{x}\ln \psi \lbrack u;\lambda ]  \label{R9}
\end{equation}%
for any $\lambda \in \mathbb{R}$ as $|\lambda |\rightarrow \infty .$ If now
to define the expression \ 
\begin{equation}
v[u;\lambda ]:=\psi \lbrack u;\lambda ]\exp (-\lambda x-{\lambda }^{n}t)\in
\Lambda ^{0}(\mathcal{\overline{K}\{}u\mathcal{\}}),  \label{R9a}
\end{equation}%
such that 
\begin{equation}
u=-D_{x}\ln v[u;\lambda ],  \label{R9b}
\end{equation}%
one can easily see from the constraint (\ref{R3}) and the equation (\ref{R6}%
) that the functional $\ \gamma _{0}=\int dxu\in \mathcal{D(}\overline{%
\mathcal{A}};dx)$ \ will be the only conserved quantity for the
differentiation $\ D_{t\text{ }}$ in the differential algebra $\overline{%
\mathcal{A}},$ if the element $\ v[u;\lambda ]\mathcal{\in }\mathcal{\ }%
\Lambda ^{0}(\mathcal{\overline{K}\{}u\mathcal{\}})$ \ satisfies a
necessarily \textit{linear} differential relationship \ 
\begin{equation}
D_{t}v[u;\lambda ]\ +A(D_{x})v[u;\lambda ]=0,  \label{R9c}
\end{equation}%
for some linear differential mapping $A(D_{x}):\Lambda ^{0}(\mathcal{%
\overline{K}\{}u\mathcal{\}})\mathcal{\rightarrow }\Lambda ^{0}(\mathcal{%
\overline{K}\{}u\mathcal{\}}),$ not depending on the element $u\in \mathcal{K%
}.$ Otherwise, \ it would then provide with an additional constraint on the
function $u\in \mathcal{K},$ giving rise to an additional \ conserved
quantity for the differentiation $\ D_{t\text{ }}$ in the reduced
differential algebra $\Lambda (\mathcal{\overline{K}\{}u\mathcal{\}}).$
Taking this reasoning into account, we can \ make the shift $\mathcal{K\ni }%
\ u\rightarrow \xi u\in \mathcal{K},\xi \in \mathbb{R},$ and consider the
relationship (\ref{R9c}) as $\xi \rightarrow 0.$ \ As a result one obtains
that the limiting function $\bar{v}[x,t;\lambda ]:=\exp (-\lambda x-\lambda
^{n}t)\in $ $\Lambda ^{0}(\mathcal{\overline{K}\{}0\mathcal{\}})$ \
satisfies the \textit{linear} differential relationship 
\begin{equation}
D_{t}\bar{v}[x,t;\lambda ]+K^{\prime ,\ast }[0]\bar{v}[x,t;\lambda ]=0,
\label{R10}
\end{equation}%
whence one derives easily that 
\begin{equation}
K^{\prime }[0]=-D_{x}^{n}  \label{R11}
\end{equation}%
for the before chosen integer $n\in \mathbb{Z}_{+}\backslash \{0,1\}.$ Thus,
from (\ref{R10}) one brings about that the following linear relationship%
\begin{equation}
D_{t}v[u;\lambda ]+(-D_{x})^{n}v[u;\lambda ]=0  \label{R12}
\end{equation}%
holds for any $u\in \mathcal{K},$ satisfying the constraint (\ref{R3}).

Turning now back to the relationship (\ref{R9b}), one can easily obtain that
the function $u\in \mathcal{K}$ \ \ under our constraint (\ref{R3})
guarantees that the only conserved quantity with respect to the
differentiation $D_{t}$ is the functional $\gamma _{0}=\int dxu\in \mathcal{%
D(}\overline{\mathcal{A}};dx)\ \ $and satisfies the following nonlinear
differential relationships:%
\begin{equation}
D_{t_{n}}u+K_{n}[u]=0\   \label{R14}
\end{equation}%
with respect to the evolution parameters $t_{n}\in \mathbb{R}$ for any $n\in 
\mathbb{Z}_{+}\backslash \{0,1\},$ where the corresponding vector fields $%
K_{n}:\mathcal{\overline{K}\{}u\}\mathcal{\rightarrow }\Gamma \mathcal{(%
\overline{K}\{}u\})$ are equal to the analytical expressions 
\begin{equation}
K_{n}[u]=\exp (\partial ^{-1}u)\sum\limits_{k=0}^{n-1}\frac{n!}{(n-k)!k!}%
(D_{x}^{n-k}u)D_{x}^{k}\exp (-\partial ^{-1}u)  \label{R14a}
\end{equation}%
of the Burgers and Korteweg-de Vries type. If to recall that $%
K_{n}[u]=D_{x}k_{n}[u],$ then one easily finds from \ (\ref{R14a}) that the
following recurrent relationship 
\begin{equation}
k_{n+1}[u]=-uk_{n}[u]+D_{x}k_{n}[u]  \label{R14b}
\end{equation}%
holds for $n\in \mathbb{N},k_{1}[u]:=u\in \mathcal{K}.$ For the simplest
cases one obtains from \ (\ref{R14b}) the constraints 
\begin{equation}
n=2:\text{ \ \ \ }D_{t_{2}}u-D_{x}^{2}u+2uD_{x}u=0,  \label{R15a}
\end{equation}%
\begin{equation}
n=3:\text{ \ \ \ }D_{t_{3}}u-D_{x}^{3}u+3D_{x}(uD_{x}u)-D_{x}(u^{3})=0\ \ 
\label{R15b}
\end{equation}%
and so on. In addition to the constraint (\ref{R14}) for $n\in \mathbb{N},$
from (\ref{R9}) one obtains the following interesting relationship on the
element $\psi \lbrack u;\lambda ]\in \Lambda ^{0}(\mathcal{\overline{K}\{}u%
\mathcal{\}}):$%
\begin{equation}
D_{x}\psi =(\lambda -u)\psi  \label{R16}
\end{equation}%
for any $\lambda \in \mathbb{R}.$ Having taken into account the condition (%
\ref{R12}), we also easily find that the linear differentiation expression 
\begin{equation}
D_{t_{n}}\psi =\sum_{k=0}^{n}\frac{n!}{(n-k)!k!}(-\lambda
)^{n-k}D_{x}^{k-1}[(\lambda -u)\psi ]  \label{R17}
\end{equation}%
holds for $n\in \mathbb{N}$ and any $\lambda \in \mathbb{R},$ which together
with the expression \ (\ref{R16}) is equivalent to the related Lax type
representations for the vector fields (\ref{R14a}). \ Thus, one can
formulate the following theorem.

\begin{theorem}
The scalar multiplicative expressions (\ref{R16}) and (\ref{R17}) are the
corresponding representations of the commuting differential relationship (%
\ref{R1}) in the differential algebra $\Lambda ^{0}(\mathcal{\overline{K}\{}u%
\mathcal{\}})\simeq \mathcal{\overline{K}\{}u\}$ under the constraints (\ref%
{R14}), being equivalent to the related Lax type representations for the
vector fields (\ref{R14a}).
\end{theorem}

Concerning the case $n=3$ some related preliminary results, including the
representations (\ref{R16}) and (\ref{R17}), were before obtained in the
work \cite{PPV,Ta}. In addition, in \cite{PPV} there was stated that the odd
members of dynamical systems \ (\ref{R17}) possess an infinite hierarchy of
conserved nonlocal quantities and two compatible \cite{Bl,Ol,PM} Poisson
structures $\ \vartheta ,\eta :T^{\ast }(\mathcal{\overline{K}}\{u\})%
\mathcal{\rightarrow }T(\mathcal{\overline{K}}\{u\})$ \ subject to the
extended ring $\mathcal{\tilde{K}}\{u\}=\mathcal{\overline{K}}\{u;\partial
^{-1}u\}$ and equal to the expressions $\ $ 
\begin{eqnarray}
\vartheta \ &=&\partial exp(\ \partial ^{-1}u)\text{ }\partial \text{ }exp(\
\partial ^{-1}u)\partial ,  \label{R18} \\
&&  \notag \\
\eta &=&\partial exp(\ \partial ^{-1}u)\text{ }\partial ^{3}\text{ }exp(\
\partial ^{-1}u)\partial .  \notag
\end{eqnarray}%
With respect to the structures (\ref{R18}) the dynamical system (\ref{R15b})
is representable as bi-Hamiltonian flow in $\ $the ring $\overline{\mathcal{K%
}}\mathcal{\{}u\}:\ $ 
\begin{equation}
du/dt_{3}=-\vartheta \func{grad}H_{3}=-\eta \func{grad}\bar{H}_{3},
\label{R19a}
\end{equation}%
where \ 
\begin{equation}
H_{3}:=\frac{1}{2}\int u^{2}\exp (-2\partial ^{-1}u)dx,\ \ \ \ \ \ \ \ \ 
\bar{H}_{3}:=\frac{1}{2}\int \exp (-2\partial ^{-1}u)dx.  \label{R19b}
\end{equation}%
Similar results can be, evidently, stated for any constraint of (\ref{R14}).$%
\ $

It is also easy to observe that the recurrent relationship (\ref{R14b}) can
be rewritten by means of the recursion operator 
\begin{equation}
\Lambda =\partial -\partial u\partial ^{-1}  \label{R20}
\end{equation}%
for the corresponding vector fields (\ref{R14a}), as%
\begin{equation}
K_{n+1}[u]=\Lambda K_{n\ }[u],  \label{R21}
\end{equation}%
$n\in \mathbb{N}.$ Moreover, as follows from (\ref{R19a}), the next
interesting algebraic relationship 
\begin{equation}
\Lambda ^{2}=\eta \vartheta ^{-1}=\partial exp(\ \partial ^{-1}u)\text{ }%
\partial ^{2}\text{ }exp(\ -\partial ^{-1}u)\partial ^{-1}  \label{R22}
\end{equation}%
holds.

\begin{corollary}
As a simple corollary from the recurrent relationship  (\ref{R21}) one
obtains that all of the vector fields (\ref{R14a}) are commuting  \cite%
{Bl,PM} to each other:%
\begin{equation}
\lbrack K_{n\ },K_{m}]=0\   \label{R23}
\end{equation}%
for any $n,m\in \mathbb{N}.$
\end{corollary}

The result above also easily follows from the differential relationship (\ref%
{R12}). Really, since its solutions for different $n\in \mathbb{Z}%
_{+}\backslash \{0,1\}$ depend only on the initial conditions $%
u|_{t_{n}=0}\in $ $\mathcal{K\{}u\},$ the corresponding differentiations 
\begin{equation}
D_{t_{n}}=\exp [(-1)^{n+1}\ D_{x}{}^{n}],  \label{R24}
\end{equation}%
being well defined on the linear Schwartz space $S(\mathbb{R};\mathbb{R}),$
are \textit{a priori }commuting to each other, that is 
\begin{equation}
\lbrack D_{t_{n}},D_{t_{m}}]=0.  \label{R25}
\end{equation}%
The latter is equaivalent to the commutation condition (\ref{R23}).

\section{\protect\bigskip Conclusion}

The linear differential relationships (\ref{R16}) and (\ref{R17}) \ in the
differential Grassmann algebra $\Lambda ^{0}(\mathcal{\overline{K}\{}u%
\mathcal{\}})\ $for $u\in \mathcal{K},$ satisfying the differential
constraint \ (\ref{R3}), realize scalar representations of the basic
differentiations $D_{t}$ and $D_{x},$ \textit{a priori} commuting to each
other on the whole functional manifold of solutions to differential
constraints (\ref{R14}). Simultaneously, they are equivalent within the well
known integrability theory interpretation as the related Lax-Zakharov-Shabat
type representation \cite{No} for the vector fields (\ref{R14a}), useful in
many cases for practical applications. It is also evident that a similar
approach can be applied to other cases, when there are imposed on the ideal $%
\mathcal{K\{}u,v,...\},$ depending on two or more different functions $%
u,v,...\in \mathcal{K},$ some \textit{a priori} given constraint $0=Z[u]\in 
\mathcal{K\{}u,v,...\}.$ In this case one can also try to construct the
corresponding finite dimensional representations of differentiations $D_{t}$
and $D_{x}$ in $\mathcal{K\{}u,v,...\}.$ For instance, the following
two-component hydrodynamic type system 
\begin{equation}
\left. 
\begin{array}{c}
u_{t}=u_{xx}+2uu_{x}+v_{x} \\ 
v_{t}=u_{x}v+uv_{x}%
\end{array}%
\right\} :=K[u,v],  \label{R26}
\end{equation}%
can be realized as a constraint imposed on the functional ring $\mathcal{K\{}%
u,v\}.\ $\ It was singled out in \cite{TW} as suspicious on the Lax type
integrability. Based on the techniques, devised in \cite{PPV,PAPP,PrY} and
in the present work \cite{POS}, we were able to prove   \cite{DPOS} the Lax
type integrability of (\ref{R26}) and to present an infinite hierarchy of
new Lax type integrable dynamical systyems of the hydrodynamic type.

\section{Acknowledgements}

Authors are gratefully acknowledge partial support of the research in this
paper from the Turkey-Ukrainian: TUBITAK-NASU Grant 110T558. \ A.K.
Prykarpatski cordially thanks Prof. A. Augustynowicz (Gdansk University,
Poland) and Prof. M.V. Pavlov (Lomonosov Sate University, Russian
Federation) for useful discussions of the results obtained. Special thanks
belong to Referee for many usefull remarks, comments and professional
suggestions which made a manuscript both more clear and readable.

\end{document}